\documentclass[reprint,showpacs,amsmath,amssymb,aps,prc]{revtex4-1}
\usepackage{bm}
\usepackage{graphicx}
\usepackage{dcolumn}
\newcommand{\be}{\begin{equation}}
\newcommand{\ee}{\end{equation}}
\newcommand{\bea}{\begin{eqnarray}}
\newcommand{\eea}{\end{eqnarray}}

\newcommand{\mbss}[1]{_{\mbox{\scriptsize #1}}}

\newcommand{\mbsu}[1]{\mbox{\scriptsize #1}}

\newcommand{\vphu}{\vphantom{*}}
\newcommand{\vphd}{\vphantom{1}}

\newcommand{\scs}{\scriptstyle}

\newcommand{\ve}{\varepsilon}

\begin{document}

\title{Subtraction method and stability condition\\
in the extended RPA theories}

\author{V. I. Tselyaev}

\affiliation{%
Nuclear Physics Department,
V. A. Fock Institute of Physics,
St. Petersburg State University, RU-198504,
St. Petersburg, Russia}

\date{\today}

\begin{abstract}
The extended RPA theories are analyzed from the point of view
of the problem of stability of their solutions.
Three kinds of such theories are considered:
the second RPA and two versions of the quasiparticle-phonon coupling
model within the time-blocking approximation:
the model including $1p1h\otimes$phonon configurations
and the two-phonon model.
It is shown that stability is ensured by making use
of the subtraction method proposed previously to solve double counting
problem in these theories.
This enables one to generalize the famous Thouless theorem proved
in the case of the RPA.
These results are illustrated by an example of schematic model.
\end{abstract}



\maketitle

\section{Introduction}
\label{sect1}

One of the trends in the modern nuclear structure theory is the calculations
in a large model configuration space. On the one hand, this trend is caused
by the requirement of the internal consistency of the theory.
On the other hand, such calculations allow us in some cases to describe
the nuclear structure effects which can not be reproduced within the framework
of the more simple models. However, the use of the large configuration space
leads to the problems of convergence and stability of the solutions obtained.
It should be noted that, though the problems of convergence and stability
are close to each other, they do not coincide.
The convergence is understood in the usual mathematical sense, while
stability, as applied to the description of the excited states,
implies that all the calculated excitation energies should be real
and positive.

The most widely used models, in which these problems can be resolved
(or do not arise at all),
are the Hartree-Fock (HF) approximation aimed at the description of
the ground states and the random phase approximation (RPA) within which
the characteristics of the excited states can be calculated (see \cite{RS80}).
Usually, these models are referred to as the mean-field theories.
In particular, the problem of stability is resolved in the HF based
self-consistent RPA as was shown by Thouless in Refs. \cite{Th60,Th61}.
But these problems become actual and remain so far open in the models
going beyond these approximations.
The reasons for developing and using such extended models are well known
(see, e.g., Ref.~\cite{DNSW90}).
First of all, they are related to the fact that within the HF approximation
and within the RPA one can not describe the effects of the fragmentation
of the nuclear states leading to the formation of the so-called
spreading widths of the resonances.

There is a series of models within which these effects
are included. One of them is the second RPA
(SRPA, see \cite{DNSW90,SW61,S62,YDG83}).
The problems mentioned above arise in this model because of enlarging
the configuration space which includes two-particles--two-holes ($2p2h$)
states in addition to the one-particle--one-hole ($1p1h$) states
incorporated in the RPA.
In Refs. \cite{PR09,PR10,GGC11,GGDDCC12} it was obtained that
calculations of giant resonances in
$^{16}$O, $^{40}$Ca, $^{48}$Ca, and $^{90}$Zr
within the SRPA lead to a very large
(up to 10~MeV and more so)
downward shifts of their centroids relatively to the RPA values
if the size of the configuration space is sufficiently large.
It was also found that some low-lying states in the SRPA become unstable,
so the question arises as to whether the SRPA is applicable in the low-energy
region (see \cite{PR10}).
In Refs. \cite{MGCG10,MGRMC12} the problem of the ultraviolet divergence
appearing at the second order beyond the HF approximation
was analyzed in the case of nuclear matter.

In the present paper we will consider in detail the problem of stability
in the extended RPA (ERPA) theories.
The problem of convergence will be only briefly touched upon.
Note that the term ERPA is sometimes used with regard to the models
taking into account the ground state correlations beyond the RPA
(see, e.g., Refs. \cite{VKCS00,TS07}). These effects will not be discussed here.
The following models will be considered:
the SRPA and two versions of the quasiparticle-phonon coupling model
formulated within the Green function method on the basis of
the time-blocking approximation (TBA):
the model including $1p1h\otimes$phonon configurations
\cite{TBA89,TBA97,KST04} and the two-phonon model \cite{QTBA1}.
It will be shown that stability can be ensured by making use
of the subtraction method \cite{QTBA1}.
This method was applied in the calculations of giant resonances
within the quasipartical TBA (QTBA) in \cite{QTBA2,ISGMR,AGKK11}
and within its relativistic generalization in \cite{RTBA,RQTBA1,RQTBA2}
to eliminate double counting in these models.
However, it was not analyzed previously in the context of the
stability issue.

The paper is organized as follows.
In Sec.~\ref{sect2} the problem of stability and
the content of the Thouless theorem in the RPA framework are considered.
In Sec.~\ref{sect3} the response function formalism is outlined within
which the stability problem is revealed in more detail.
The ERPA theories mentioned above are briefly described in Sec.~\ref{sect4}.
In Sections \ref{sect5} and \ref{sect6} the subtraction method and
the stability condition in the ERPA are considered.
The general results obtained in the previous sections are analyzed
within the framework of the schematic model in Sec.~\ref{sect7}.
The conclusions are given in Sec.~\ref{sect8}.

\section{Thouless theorem}
\label{sect2}

The Thouless theorem \cite{Th60,Th61} determines stability condition
in the case of the self-consistent RPA.
Let us briefly recall the structure of the RPA equations and
the content of this theorem because its generalization to the case
of the ERPA theories can be carried out (see Sec.~\ref{sect6})
using a simple analogy.

To build the RPA equations one needs
the single-particle density matrix $\rho^{\vphu}_{12}$,
the single-particle Hamiltonian $h^{\vphu}_{12}$, and
the amplitude of the residual interaction ${V}^{\vphu}_{12,34}$.
Here and in the following the numerical indices ($1,2,3,\ldots$)
stand for the sets of the quantum numbers of some single-particle basis.
It is supposed that the following equalities are fulfilled
\be
\rho^2=\rho\,,\qquad [\,h,\rho\,]=0\,.
\label{speqs}
\ee
Let us introduce the single-particle basis that diagonalizes operators
$h$ and $\rho\,$:
\be
h^{\vphu}_{12} = \ve^{\vphu}_{1}\delta^{\vphu}_{12}\,,
\qquad
\rho^{\vphu}_{12} = n^{\vphu}_{1}\delta^{\vphu}_{12}
\label{spbas}
\ee
where $n^{\vphu}_{1}$ is the occupation number.
In what follows the indices $p$ and $h$ will be used to label
the single-particle states of the particles ($n^{\vphu}_{p} = 0$)
and holes ($n^{\vphu}_{h} = 1$) in this basis.
The matrix
\be
M^{\mbss{RPA}}_{12,34} =
\delta^{\vphu}_{13}\,\rho^{\vphu}_{42} -
\rho^{\vphu}_{13}\,\delta^{\vphu}_{42}
\label{mrpa}
\ee
is the metric matrix in the RPA.
The range of $M^{\mbss{RPA}}_{\vphd}$ forms the $1p1h$ configuration space.
The vectors $z^{\vphu}_{12}$ in this space have the components
of $z^{\vphu}_{ph}$ and $z^{\vphu}_{hp}$ types.
The RPA matrix $\Omega^{\mbss{RPA}}_{12,34}$ acts in the $1p1h$ space.
In the general case it has the form
\be
\Omega^{\mbss{RPA}}_{12,34} =
h^{\vphu}_{13}\,\delta^{\vphu}_{42} -
\delta^{\vphu}_{13}\,h^{\vphu}_{42} + \sum_{56}
M^{\mbss{RPA}}_{12,56}\,{V}^{\vphu}_{56,34}\,.
\label{orpa}
\ee
The RPA eigenvalue equation reads
\be
\sum_{34} \Omega^{\mbss{RPA}}_{12,34}\,z^{n}_{34} =
\omega^{\vphu}_n\,z^{n}_{12}
\label{rpabe}
\ee
where $\omega^{\vphu}_n$ is the excitation energy,
$z^{n}_{12}$ is the transition amplitude.
In the case of the self-consistent RPA, based on
the energy density functional $E[\rho]$, the quantities $h$ and $V$
in Eq.~(\ref{orpa}) are linked by the equations
\be
h^{\vphu}_{12} = \frac{\delta E[\rho]}{\delta\rho^{\vphu}_{21}}\,,
\qquad
{V}^{\vphu}_{12,34} =
\frac{\delta^2 E[\rho]}
{\delta\rho^{\vphu}_{21}\,\delta\rho^{\vphu}_{34}}\,.
\label{frpa}
\ee
Eqs. (\ref{speqs}) play the role of the equations of motion.

The Thouless theorem can be formulated in terms of the following
general statement (see, e.g., Ref.~\cite{RS80}).
Let a matrix $A$ be representable in the form $A=BC$ where the matrices
$B$ and $C$ are Hermitian and $C$ is positive semidefinite
(i.e., $\,\langle\,z\,|\,C\,|\,z\,\rangle \geqslant 0\,$
for any complex vector $|\,z\,\rangle$).
Then all the eigenvalues of the matrix $A$ are real.
Indeed, consider the eigenvalue equation
\be
A\,|\,x\,\rangle = a\,|\,x\,\rangle\,.
\label{aevq}
\ee
From the positive semidefiniteness of the Hermitian matrix $C$
it follows that there exists Hermitian matrix $C^{1/2}$ such that
$C=(C^{1/2})^2$. Let us denote $\,|\,y\,\rangle=C^{1/2}|\,x\,\rangle$.
If $\,|\,y\,\rangle=0$ then $a=0$. If $\,|\,y\,\rangle \ne 0$
then, by multiplying Eq.~(\ref{aevq}) with $C^{1/2}$, we obtain
$D\,|\,y\,\rangle = a\,|\,y\,\rangle$ where $D=C^{1/2}B\,C^{1/2}$.
The matrix $D$ is Hermitian, consequently, the eigenvalue $a$ is real.

Coming back to Eq.~(\ref{rpabe}),
let us define the RPA stability matrix
\be
\mathfrak{S}^{\mbss{RPA}}_{\vphd} =
M^{\mbss{RPA}}_{\vphd}\Omega^{\mbss{RPA}}_{\vphd}.
\label{srpa}
\ee
Since $(M^{\mbss{RPA}}_{\vphd})^2=1$ in the $1p1h$ space,
Eq.~(\ref{srpa}) is equivalent to the equation
\be
\Omega^{\mbss{RPA}}_{\vphd}=M^{\mbss{RPA}}_{\vphd}
\mathfrak{S}^{\mbss{RPA}}_{\vphd}.
\label{omsrpa}
\ee
Now we note that
both the matrix $M^{\mbss{RPA}}_{\vphd}$ and the matrix
$\mathfrak{S}^{\mbss{RPA}}_{\vphd}$ in Eq.~(\ref{omsrpa}) are Hermitian.
Therefore, all eigenvalues $\omega^{\vphu}_n$ in Eq. (\ref{rpabe})
are real if the stability matrix $\mathfrak{S}^{\mbss{RPA}}_{\vphd}$
is positive semidefinite. This is the statement of the Thouless theorem.
The positive semidefiniteness of the matrix
$\mathfrak{S}^{\mbss{RPA}}_{\vphd}$ follows from the conditions of
minimization of the energy density functional $E[\rho]$
in the self-consistent theory (see \cite{RS80,Th60}).
Note that the matrix $\mathfrak{S}^{\mbss{RPA}}_{\vphd}$
is not positive definite because of the symmetry properties of $E[\rho]$.

Reality of the eigenvalues in Eq.~(\ref{rpabe}) leads to the
following symmetry property of the solutions of this equation.
Let us introduce the permutation operator acting in the space of the
pairs of the single-particle indices:
$\mathfrak{P}^{\vphu}_{12,34}=\delta^{\vphu}_{14}\delta^{\vphu}_{23}\,$.
From the definitions (\ref{mrpa}), (\ref{orpa}), and (\ref{frpa})
it follows that $\Omega^{\mbss{RPA}}_{\vphd}=
-\mathfrak{P}\,\Omega^{\mbss{RPA}*}_{\vphd}\mathfrak{P}$.
This equality together with Eq.~(\ref{rpabe})
and reality of $\omega^{\vphu}_n$ brings us to the equation
\be
|\,z^{-n}\rangle = \mathfrak{P}\,|\,z^{n}\rangle^*
\label{simz1}
\ee
where the eigenvectors $|\,z^{n}\rangle$ and $|\,z^{-n}\rangle$
correspond to the eigenvalues $\omega^{\vphu}_n$ and $-\omega^{\vphu}_n$,
respectively.

\section{Response function formalism}
\label{sect3}

The other important consequences of the positive semidefiniteness
of the stability matrix which will be used in the following
in the context of the ERPA theories
concern the properties of the response function $R(\omega)$
defined in the RPA by the equation
\be
R^{\mbss{RPA}}_{\vphd}(\omega) = -
\bigl(\,\omega - \Omega^{\mbss{RPA}}_{\vphd}\bigr)^{-1}
M^{\mbss{RPA}}_{\vphd}.
\label{rfdef1}
\ee
An overall sign in this formula is chosen in accordance with
the usual definition of the response function in the Green function
method (see Ref.~\cite{SWW77}).
The response function formalism is a conventional tool for the description
of nuclear excitations. In the general case
the distribution of the strength of transitions
in the nucleus caused by some external field represented by the
single-particle operator $Q$ is determined by the (dynamic)
polarizability $\Pi(\omega)$ which is defined in terms of the response
function as
\be
\Pi(\omega) = - \langle\,Q\,|\,R(\omega)\,|\,Q\,\rangle\,.
\label{poldef}
\ee
The poles and residua of the function $\Pi(\omega)$ coincide with
the excitation energies and
the transition probabilities (see Eq.~(\ref{piqrpa}) below).

Let us introduce an auxiliary matrix
\be
\tilde{\mathfrak{S}}^{\mbss{RPA}}_{\vphd}=
\mathfrak{S}^{\mbss{RPA}}_{\vphd}+\delta
\label{strpa}
\ee
where $\delta$ is a real positive number.
If $\mathfrak{S}^{\mbss{RPA}}_{\vphd}$ is positive semidefinite, then
the matrix $\tilde{\mathfrak{S}}^{\mbss{RPA}}_{\vphd}$ is positive definite
and consequently there exists invertible Hermitian matrix $\tilde{\mathfrak{S}}^{1/2}$
such that $\tilde{\mathfrak{S}}^{\mbss{RPA}}_{\vphd}=(\tilde{\mathfrak{S}}^{1/2})^2$.
Let us denote:
$\tilde{\Omega}^{\mbss{RPA}}_{\vphd}=M^{\mbss{RPA}}_{\vphd}
\tilde{\mathfrak{S}}^{\mbss{RPA}}_{\vphd}$,
\be
\tilde{R}^{\mbss{RPA}}_{\vphd}(\omega) = -
\bigl(\,\omega - \tilde{\Omega}^{\mbss{RPA}}_{\vphd}\bigr)^{-1}
M^{\mbss{RPA}}_{\vphd}.
\label{rfdef1t}
\ee
Using the invertibility of the matrix $\tilde{\mathfrak{S}}^{1/2}$ we obtain
\bea
&&\tilde{R}^{\mbss{RPA}}_{\vphd}(\omega)
\nonumber\\
&& = - (\tilde{\mathfrak{S}}^{1/2})^{-1}
\bigl(\,\omega - \tilde{H}^{\mbss{RPA}}_{\vphd}\bigr)^{-1}
\tilde{H}^{\mbss{RPA}}_{\vphd}\,(\tilde{\mathfrak{S}}^{1/2})^{-1}
\label{rfhrpa}
\eea
where
$\tilde{H}^{\mbss{RPA}}_{\vphd} = \tilde{\mathfrak{S}}^{1/2}
M^{\mbss{RPA}}_{\vphd}\tilde{\mathfrak{S}}^{1/2}$.
The matrices $\tilde{\Omega}^{\mbss{RPA}}_{\vphd}$ and
$\tilde{H}^{\mbss{RPA}}_{\vphd}$ have the same set
of the (non-zero) eigenvalues $\{\,\tilde{\omega}^{\vphu}_{n} \}$.
But, in contrast to $\tilde{\Omega}^{\mbss{RPA}}_{\vphd}$,
the matrix $\tilde{H}^{\mbss{RPA}}_{\vphd}$ is Hermitian.

Let $\{\,|\,\tilde{y}^{n} \rangle \}$ be a complete set
of the orthonormalized eigenvectors of the matrix
$\tilde{H}^{\mbss{RPA}}_{\vphd}$.
Insertion of the sum
$\sum_{n}|\,\tilde{y}^{n} \rangle \langle\,\tilde{y}^{n}|=1$
into Eq. (\ref{rfhrpa}) yields
\be
\tilde{R}^{\mbss{RPA}}_{\vphd}(\omega) = -
\sum_{n} \frac{\mbox{sgn}(\tilde{\omega}^{\vphu}_{n})\,
|\,\tilde{z}^{n} \rangle \langle \tilde{z}^{n}|}
{\omega - \tilde{\omega}^{\vphu}_{n}}
\label{rfexp2}
\ee
where
\bea
|\,\tilde{z}^{n} \rangle
&=& \sqrt{|\,\tilde{\omega}^{\vphu}_n|}
\,(\tilde{\mathfrak{S}}^{1/2})^{-1}|\,\tilde{y}^{n} \rangle
\nonumber\\
&=& \frac{\mbox{sgn}(\tilde{\omega}^{\vphu}_{n})}
{\sqrt{|\,\tilde{\omega}^{\vphu}_n}|}\;M^{\mbss{RPA}}_{\vphd}
\,\tilde{\mathfrak{S}}^{1/2}\,|\,\tilde{y}^{n} \rangle\,,
\label{zydef}
\eea
\be
\tilde{\Omega}^{\mbss{RPA}}_{\vphd}|\,\tilde{z}^{n} \rangle
= \tilde{\omega}^{\vphu}_n\,|\,\tilde{z}^{n} \rangle\,,
\label{trpae}
\ee
\be
\langle\,\tilde{z}^{n}\,|\,M^{\mbss{RPA}}_{\vphd}
|\,\tilde{z}^{n'} \rangle =
\mbox{sgn}(\tilde{\omega}^{\vphu}_{n})\,\delta^{\vphu}_{n,\,n'}\,.
\label{tzmz}
\ee
Now, going to the limit $\delta \rightarrow +0$, we obtain
\be
R^{\,\mbsu{RPA}}_{\vphd}(\omega) =
R^{\,\mbsu{RPA(0)}}(\omega)
- {\sum_{n}}^{\,\prime}
\frac{\;\mbox{sgn}(\omega^{\vphu}_{n})\,a^{n}}
{\omega - \omega^{\vphu}_n}
\label{rfexp3}
\ee
where
\be
R^{\,\mbsu{RPA(0)}}(\omega) = - \sum_{k = 1}^2
\,\frac{a^{(\,0,k)}}{\omega^k}\,,
\label{rfrpa0}
\ee
\be
a^{(\,0,k)} = \lim_{\delta \rightarrow +0}\;
{\sum_{n}}^{(0)}\mbox{sgn}(\tilde{\omega}^{\vphu}_{n})\,
\tilde{\omega}^{k-1}_n\,
|\,\tilde{z}^{n} \rangle \langle \tilde{z}^{n}|\,,
\label{bkdef}
\ee
\be
a^{n} =
|\,z^{n} \rangle \langle z^{n}|\,.
\label{ansdef}
\ee
Symbol ${\sum}^{(0)}$ in Eq.~(\ref{bkdef})
means the sum over all the states $n$ for which
$\tilde{\omega}^{\vphu}_n \rightarrow \pm 0$ at
$\delta \rightarrow +0$
(that is over the spurious states).
Symbol ${\sum}^{\,\prime}$ in Eq.~(\ref{rfexp3}) means the sum
over all the states $n$ excluding the spurious states.
Note that the sum over $k$ in Eq.~(\ref{rfrpa0}) is limited to the
first two terms because,
as follows from Eqs. (\ref{zydef}) and (\ref{bkdef}),
$a^{(\,0,k)}=0$ at $k>2$.

The non-spurious eigenvectors $|\,z^{n} \rangle$
satisfy Eq.~(\ref{rpabe}) and are normalized by the condition
\be
\langle\,{z}^{n}\,|\,M^{\mbss{RPA}}_{\vphd}
|\,{z}^{n'} \rangle =
\mbox{sgn}(\omega^{\vphu}_{n})\,\delta^{\vphu}_{n,\,n'}
\label{zmz}
\ee
following from Eq. (\ref{tzmz}).
The matrices $a^{(\,0,1)}$ and $a^{(\,0,2)}$ are Hermitian and
satisfy the equations
\be
\Omega^{\mbss{RPA}}_{\vphd}a^{(\,0,1)} = a^{(\,0,2)},\quad
\Omega^{\mbss{RPA}}_{\vphd}a^{(\,0,2)} = 0\,,
\label{b1b2e1}
\ee
\be
a^{(\,0,1)}M^{\mbss{RPA}}_{\vphd}\,a^{(\,0,k)} = a^{(\,0,k)},
\label{b1b2e2}
\ee
\be
a^{(\,0,1)} = -\mathfrak{P}\,a^{(\,0,1)*}\mathfrak{P},\quad
a^{(\,0,2)} =  \mathfrak{P}\,a^{(\,0,2)*}\mathfrak{P}
\label{b1b2e3}
\ee
following from Eqs. (\ref{simz1}), (\ref{trpae}), (\ref{tzmz}),
and (\ref{bkdef}). The closure relation
\be
a^{(\,0,1)} + {\sum_{n}}^{\,\prime}
\mbox{sgn}(\omega^{\vphu}_{n})\,a^{n} = M^{\mbss{RPA}}_{\vphd}
\label{rpacr}
\ee
follows from Eqs. (\ref{rfdef1}), (\ref{rfexp3}), and (\ref{rfrpa0}).

Eq.~(\ref{ansdef}) implies that all the matrices $a^{n}$
in the expansion (\ref{rfexp3}) are Hermitian and positive semidefinite.
In addition from Eqs. (\ref{simz1}) and (\ref{ansdef}) we get
\be
a^{-n} = \mathfrak{P}\;a^{n\,*}\,\mathfrak{P}\,.
\label{simans}
\ee
These properties of the residua
of the function $R^{\mbss{RPA}}_{\vphd}(\omega)$
coincide with the properties of the exact response function
following from its spectral representation
(see, e.g., Ref.~\cite{RS80}).
Taking this into account and making use of Eq.~(\ref{poldef})
we obtain that
\be
\Pi^{\mbss{RPA}}_{\vphd}(\omega) =
{\sum_{n}}^{\,\prime}
\frac{\;\mbox{sgn}(\omega^{\vphu}_{n}) B^{\vphu}_{\,n}(Q)}
{\omega - \omega^{\vphu}_n}
\label{piqrpa}
\ee
where transition probabilities
$B^{\vphu}_{\,n}(Q) = \langle\,Q\,|\,a^{n}\,|\,Q\,\rangle$
are real and non-negative and it is supposed that
\be
\langle\,Q\,|\,a^{(\,0,k)}\,|\,Q\,\rangle = 0\,,\quad k=1,2\,.
\label{qa0q}
\ee
From Eq.~(\ref{simans}) we also obtain that
$B^{\vphu}_{-n}(Q) = B^{\vphu}_{\,n}(Q^{\dag})$.

If the stability matrix does not possess the property of
the positive semidefiniteness,
the reality of the RPA eigenvalues $\omega^{\vphu}_n$ and
the Hermiticity and the positive semidefiniteness
of the matrices $a^{n}$ are not guaranteed.
In particular, this means that the eigenvectors with positive eigenvalues
may have negative norms.
As a consequence, the reality and the non-negativeness
of the RPA transition probabilities $B^{\vphu}_{\,n}(Q)$
in Eq.~(\ref{piqrpa}) is also not guaranteed, and the strength function
\be
S^{\mbss{RPA}}_{\vphd}(E,\Delta)=-\frac{1}{\pi}\;\mbox{Im}\,
\Pi^{\mbss{RPA}}_{\vphd}(E+i\Delta)
\label{sfdef}
\ee
may take negative values at $E>0$ and $\Delta>0$.
Note that the problem of the ``negative transition probabilities''
arising in this case can be treated as the problem of the
``negative energies'' since the function
$\Pi^{\mbss{RPA}}_{\vphd}(\omega)$ will have positive residua
at the poles $\omega=\omega^{\vphu}_n<0$.

\section{Extended RPA}
\label{sect4}

In the ERPA theories the eigenvalue equation (\ref{rpabe})
is usually replaced (see, e.g., \cite{DNSW90}) by the equation
with the energy-dependent matrix
$\Omega^{\mbss{ERPA}}_{\vphd}(\omega)$:
\be
\sum_{34} \Omega^{\mbss{ERPA}}_{12,34}
(\omega^{\vphu}_{\nu})\,z^{\nu}_{34} =
\omega^{\vphu}_{\nu}\,z^{\nu}_{12}
\label{erpabe}
\ee
where $\Omega^{\mbss{ERPA}}_{\vphd}(\omega)$
can be represented in the form
\be
\Omega^{\mbss{ERPA}}_{\vphd}(\omega) =
\Omega^{\mbss{RPA}}_{\vphd} +
M^{\mbss{RPA}}_{\vphd}\,{W}(\omega)
\label{oerpa1}
\ee
and it is supposed that Eqs. (\ref{erpabe}) and (\ref{oerpa1})
are written in the $1p1h$ subspace.
The matrix ${W}(\omega)$ is the interaction amplitude that includes
contributions of complex ($2p2h$) configurations.
It has the following generic form
\be
{W}(\omega) = {F}
(\,\omega - M^{\mbss{C}}_{\vphd}{\mathfrak{S}}^{\mbss{C}}_{\vphd}\,)^{-1}
M^{\mbss{C}}_{\vphd}{F}^{\dag}
\label{wdef1}
\ee
where ${\mathfrak{S}}^{\mbsu{C}}$, $M^{\mbss{C}}_{\vphd}$,
and ${F}$ are the block matrices of the form
\be
{\mathfrak{S}}^{\mbss{C}}_{\vphd} =
\left(
\begin{array}{cc}
{\mathfrak{S}}^{\mbsu{C}(+)} & 0 \\ 0 & {\mathfrak{S}}^{\mbsu{C}(-)} \\
\end{array}
\right),
\quad
M^{\mbss{C}}_{\vphd} =
\left(
\begin{array}{rr}
1 & 0 \\ 0 & -1 \\
\end{array}
\right),
\label{hcmcd}
\ee
${F}=\bigl(\,{F}^{(+)},{F}^{(-)}\bigr)$.
The Hermitian matrices ${\mathfrak{S}}^{\mbsu{C}(\pm)}$,
${\mathfrak{S}}^{\mbsu{C}}$, and $M^{\mbss{C}}_{\vphd}$
act in the subspace of complex configurations.
The matrices ${F}$ and ${F}^{\dag}$ connect this subspace
with the $1p1h$ subspace.
The matrices ${\mathfrak{S}}^{\mbsu{C}}$ and $M^{\mbsu{C}}$ play the role
of the stability matrix and the metric matrix in the $2p2h$ subspace,
respectively. In addition, the following equalities are fulfilled
\be
{\mathfrak{S}}^{\mbsu{C}(-)} = {\mathfrak{S}}^{\mbsu{C}(+)*},\quad
{F}^{(-)} = \mathfrak{P}\,{F}^{(+)*},
\label{fhsym}
\ee
\be
M^{\mbss{RPA}}_{\vphd}{F}^{(\pm)} = \pm\,{F}^{(\pm)}.
\label{fsym1}
\ee
Eqs. (\ref{fhsym}) lead to the symmetry property
\be
\Omega^{\mbss{ERPA}}_{\vphd}(\omega) =
- \mathfrak{P}\,\Omega^{\mbss{ERPA}*}_{\vphd}(-\omega^*)\mathfrak{P}
\label{oerpasym}
\ee
from which we obtain for the eigenvectors with the real eigenvalues
in Eq.~(\ref{erpabe}) the following relation
\be
|\,z^{-\nu}\rangle =
\mathfrak{P}\,|\,z^{\nu}\rangle^*,
\label{simz2}
\ee
where $|\,z^{-\nu}\rangle$ is the eigenvector with the eigenvalue
$-\omega^{\vphu}_{\nu}\,$,
as in the case of the RPA, see Eq.~(\ref{simz1}).

Using the complete sets of the eigenvectors of the
matrices ${\mathfrak{S}}^{\mbsu{C}(\pm)}$ one can represent Eq.~(\ref{wdef1})
in the more explicit form:
\be
{W}^{\vphu}_{12,34}(\omega) =
\sum_{c,\;\sigma}\,\frac{\sigma\,
{F}^{c(\sigma)}_{12}
{F}^{c(\sigma)*}_{34}}
{\omega - \sigma\,\Omega^{\vphu}_{c}}
\label{wdef2}
\ee
where $\sigma = \pm 1$, $c$ is an index of the subspace of complex
configurations, $\Omega^{\vphu}_{c}$ are the eigenvalues of the matrices
${\mathfrak{S}}^{\mbsu{C}(\pm)}$.
It is supposed that the matrices ${\mathfrak{S}}^{\mbsu{C}(\pm)}$
are positive definite and, consequently, $\Omega^{\vphu}_{c} > 0$.
Consider three models which can be formulated using Eq.~(\ref{wdef2})
for the matrix ${W}(\omega)$.
From Eqs. (\ref{fhsym}) and (\ref{fsym1}) it follows that
${F}^{c(-)}_{12}={F}^{c(+)*}_{21}$ and
${F}^{c(-)}_{ph}={F}^{c(+)}_{hp}=0$.
So only the quantities ${F}^{c(+)}_{ph}$ and the energies
$\Omega^{\vphu}_{c}$ should be specified.

(a) Second RPA in the so-called diagonal approximation
\cite{DNSW90,YDG83,PR09,PR10,GGC11,GGDDCC12}.
In this case one can set $c = \{p',p'',h',h''\}$,
\be
\Omega^{\vphu}_{c} =
\ve^{\vphu}_{p'} + \ve^{\vphu}_{p''} -
\ve^{\vphu}_{h'} - \ve^{\vphu}_{h''}\,,
\label{odef0}
\ee
\bea
{F}^{c(+)}_{ph} &=& \frac{1}{2}\;\biggl(\,
\delta^{\vphu}_{pp'}\,w^{\vphu}_{h'h'',\;p''h} -
\delta^{\vphu}_{pp''}\,w^{\vphu}_{h'h'',\;p'h}
\nonumber\\
&+&
\delta^{\vphu}_{hh'}\,w^{\vphu}_{ph'',\;p'p''} -
\delta^{\vphu}_{hh''}\,w^{\vphu}_{ph',\;p'p''}\,\biggr)
\label{fdef0}
\eea
where $w^{\vphu}_{12,34}=-w^{\vphu}_{21,34}=-w^{\vphu}_{12,43}$
is an antisymmetrized amplitude of the two-particle interaction.
In the first order we have ${V}^{\vphu}_{12,34}=w^{\vphu}_{14,23}\,$,
however in the general case the amplitude of the residual interaction
in Eq.~(\ref{frpa}) does not coincide with $w$ and is not antisymmetric.
Note that the full SRPA scheme is usually formulated by means
of the equations similar to Eq.~(\ref{wdef1}).

(b) TBA1: the quasiparticle-phonon coupling model within the TBA
including $1p1h\otimes$phonon configurations \cite{TBA89,TBA97,KST04}
(without ground state correlations beyond the RPA included in
\cite{TBA89,TBA97,KST04}).
In this case $c = \{p',h',n\}$ where $n$ is the phonon's index,
\be
\Omega^{\vphu}_{c} =
\ve^{\vphu}_{p'} - \ve^{\vphu}_{h'} + \omega^{\vphu}_{n}\,,
\quad \omega^{\vphu}_{n}>0\,,
\label{odef1}
\ee
\be
{F}^{c(+)}_{ph} =
\delta^{\vphu}_{pp'}\,g^{n}_{h'h} -
\delta^{\vphu}_{h'h}\,g^{n}_{pp'},
\label{fdef1}
\ee
$g^{n}_{12}$ is an amplitude of the quasiparticle-phonon interaction.
In the self-consistent approach,
these amplitudes (along with the phonon's energies $\omega^{\vphu}_{n}$)
are determined by the positive frequency solutions of the RPA equations
(\ref{rpabe}) and (\ref{zmz}) according to the formula
\be
g^{n}_{12} = \sum_{34} {V}^{\vphu}_{12,34}\,z^{n}_{34}.
\label{gdef}
\ee
Physical effects taken into account in the TBA1 and the general
structure of the equations are the same
as in the particle-vibration coupling model \cite{W85} and
in the model of the coupling of $1p1h$ configurations to the
doorway states \cite{CGBB94}.

(c) TBA2: the quasiparticle-phonon coupling model within the TBA
including two-phonon configurations \cite{QTBA1}.
This model is a straightforward generalization of the TBA1
by including additional correlations between particles and holes
entering $1p1h\otimes$phonon configurations
(but also without ground state correlations beyond the RPA and
without pairing correlations included in \cite{QTBA1}).
Physically, this is similar, but not equivalent in details,
to the first versions of the Quasiparticle-Phonon Model \cite{S92}.
Relativistic extension of the two-phonon model was developed
in Ref.~\cite{RQTBA2}. In the TBA2 we have: $c = \{n,n'\}$
where $n$ and $n'$ are the phonon's indices,
\be
\Omega^{\vphu}_{c} =
\omega^{\vphu}_{n} + \omega^{\vphu}_{n'}\,,
\quad \omega^{\vphu}_{n}>0\,,\quad \omega^{\vphu}_{n'}>0\,,
\label{odef2}
\ee
\be
{F}^{c(+)}_{ph} = \sum_{p''h''}\bigl(\,
\delta^{\vphu}_{pp''}\,g^{n}_{h''h} -
\delta^{\vphu}_{h''h}\,g^{n}_{pp''}\,\bigr)\,z^{n'}_{p''h''}\,.
\label{fdef2}
\ee
The amplitudes $g^{n}_{12}$, $z^{n'}_{12}$ and the phonon's
energies are determined by Eqs. (\ref{rpabe}), (\ref{zmz}),
and (\ref{gdef}), as in the TBA1.

Obviously, the TBA2 reduces to the TBA1 in the case when the second
phonon is non-collective, i.e., when
$\omega^{\vphu}_{n'}=\ve^{\vphu}_{p'}-\ve^{\vphu}_{h'}$
in Eq.~(\ref{odef2}) and
$z^{n'}_{p''h''}=\delta^{\vphu}_{p''p'}\,\delta^{\vphu}_{h''h'}$
in Eq.~(\ref{fdef2}).
However, the connection between the TBA1 and the SRPA is not so simple
because of the well-known problem of the second order contributions
arising in the quasiparticle-phonon coupling model
(see, e.g., Ref.~\cite{MBBD85}).

\section{Subtraction method}
\label{sect5}

The starting point of the ERPA theories is the usual RPA.
In many practically significant cases (except for the so-called
{\it ab initio} approaches) the self-consistent RPA is based on
the density functional theory
(DFT, see,  e.g., Refs. \cite{E03,BHR03,DFP10})
in which the energy density functional $E[\rho]$
is constructed in such a way as to provide an optimal
(exact in the limiting case) description
of the nuclear ground-state properties. Therefore, $E[\rho]$
already effectively contains a part of the contributions of those
complex configurations which are explicitly included in the ERPA.
This part can be treated as the static contributions,
in contrast to the dynamic ones which lead to the formation of
the spreading widths of the resonances. To avoid double counting,
these static contributions in the ERPA should be eliminated.
A simple way to do this is to impose the condition:
\be
\Omega^{\mbss{ERPA}}_{\vphd}(0) =
\Omega^{\mbss{RPA}}_{\vphd}\,.
\label{erpac}
\ee

The reasons for this condition are as follows.
Let $Q$ be a local Hermitian single-particle operator representing
some external field. Dynamic polarizability $\Pi(\omega)$
corresponding to this field is defined by Eq.~(\ref{poldef})
in which the response function $R(\omega)$ is defined by
Eq.~(\ref{rfdef1}) in the RPA and by the equation
\be
R^{\mbss{ERPA}}_{\vphd}(\omega) = -
\bigl(\,\omega - \Omega^{\mbss{ERPA}}_{\vphd}(\omega)\,\bigr)^{-1}
M^{\mbss{RPA}}_{\vphd}
\label{rfdef2}
\ee
in the ERPA.
Consider an energy density functional
\be
{\cal E}[\rho,\lambda]=E[\rho]+\lambda\,\mbox{Tr}\,\bigl(\rho\,Q\bigr)
\ee
where $\lambda$ is a real parameter.
According to the so-called dielectric theorem \cite{BLM79}, we have:
\be
\Pi^{\mbss{RPA}}_{\vphd}(0) = -2m^{\mbss{RPA}}_{-1} =
\biggl(\frac{d}{d\lambda}\mbox{Tr}\,\bigl(\rho^{(\lambda)}Q\bigr)\biggr)
_{\lambda = 0}
\label{dielth}
\ee
where $\Pi^{\mbss{RPA}}_{\vphd}(0)$
is the (static) polarizability calculated by making use of
Eqs. (\ref{mrpa}), (\ref{orpa}), (\ref{frpa}),
(\ref{rfdef1}), and (\ref{poldef}) in the self-consistent
RPA based on the functional $E[\rho]$, the quantity
$m^{\mbss{RPA}}_{-1}$ is the inverse energy-weighted moment
of the strength distribution in the RPA,
$\rho^{(\lambda)}$ is the equilibrium density matrix of the functional
${\cal E}[\rho,\lambda]$, and it is supposed that
$a^{(\,0,1)}|\,Q\,\rangle = 0$.
Assuming, in accordance with general principles of the DFT,
that this theory gives in a sense an exact value of the quantity
$\mbox{Tr}\,\bigl(\rho^{(\lambda)}Q\bigr)$
at any $\lambda$ near the point $\lambda=0$, one can consider that
$\Pi^{\mbss{RPA}}_{\vphd}(0)=-2m^{\vphu}_{-1}$
where $m^{\vphu}_{-1}$ is the exact inverse energy-weighted moment
of the strength distribution including contributions of all configurations.
Then, the condition
$\Pi^{\mbss{ERPA}}_{\vphd}(0)=
\Pi^{\mbss{RPA}}_{\vphd}(0)$ is natural and from this,
using Eqs. (\ref{rfdef1}), (\ref{poldef}), and (\ref{rfdef2}),
we arrive at the condition (\ref{erpac}).
This condition will be fulfilled if we change the definition of the matrix
$\Omega^{\mbss{ERPA}}_{\vphd}(\omega)$
taking instead of Eq.~(\ref{oerpa1}) the following ansatz
\be
\Omega^{\mbss{ERPA}}_{\vphd}(\omega) =
\Omega^{\mbss{RPA}}_{\vphd} + M^{\mbss{RPA}}_{\vphd}
\bigl[\,{W}(\omega) -\kappa\,{W}(0)\,\bigr]
\label{oerpa2}
\ee
and setting $\kappa=1$.

Thus, the method of eliminating the double counting consists in subtracting
the static part ${W}(0)$ from the interaction amplitude ${W}(\omega)$
containing the contributions of complex configurations.
This method was used in the calculations of giant resonances
both within self-consistent \cite{ISGMR,AGKK11,RTBA,RQTBA1,RQTBA2}
and within non-self-consistent \cite{QTBA2} approaches.
In the non-self-consistent models the problem of
double counting arises because of the use of the phenomenologically
fitted mean field and the residual interaction. In this case
the subtraction method plays the same role as the so-called refinement
procedure applied in Refs. \cite{DNSW90,TBA89,TBA97,KST04}.

\section{Stability condition in the extended RPA theories}
\label{sect6}

To analyze the properties of Eq.~(\ref{erpabe}) with the matrix
$\Omega^{\mbss{ERPA}}_{\vphd}(\omega)$ defined by
Eqs. (\ref{oerpa2}), (\ref{wdef1}), and (\ref{hcmcd})
let us recast Eq.~(\ref{erpabe}) in the extended space
including $1p1h$ and complex ($2p2h$) configurations.
Let us define the energy-independent matrix
$\widehat{\Omega}^{\mbss{ERPA}}_{\vphd}$ in this space
as the block matrix of the following form
\be
\widehat{\Omega}^{\mbss{ERPA}}_{\vphd} =
\left(
\begin{array}{cc}
\Omega^{\mbsu{RPA}(\kappa)} &
\;M^{\mbss{RPA}}_{\vphd}{F} \\
M^{\mbss{C}}_{\vphd}{F}^{\dag} &
\;M^{\mbss{C}}_{\vphd}{\mathfrak{S}}^{\mbss{C}}_{\vphd} \\
\end{array}
\right)
\label{oerpa3}
\ee
where
\be
\Omega^{\mbsu{RPA}(\kappa)} = \Omega^{\mbss{RPA}}_{\vphd}
+ \kappa\,M^{\mbss{RPA}}_{\vphd}{F}\,({\mathfrak{S}}^{\mbss{C}}_{\vphd})^{-1}
{F}^{\dag}.
\label{orpa2}
\ee
It is easy to verify that Eq.~(\ref{erpabe}) is equivalent
to the following linear eigenvalue equation
\be
\widehat{\Omega}^{\mbss{ERPA}}_{\vphd}\,|\,Z^{\,\nu}_{\vphd} \rangle  =
\omega^{\vphu}_{\nu}\,|\,Z^{\,\nu}_{\vphd} \rangle
\label{erpabe2}
\ee
where
\be
|\,Z^{\,\nu}_{\vphd} \rangle =
\left(
\begin{array}{c}
|\,z^{\,\nu}_{\vphd} \rangle \vphantom{\frac{A}{B}} \\
|\,\zeta^{\,\nu}_{\vphd} \rangle \vphantom{\frac{A}{B}} \\
\end{array}
\right).
\label{znext}
\ee
The vector $|\,z^{\,\nu}_{\vphd} \rangle$ in Eq.~(\ref{znext}) belongs
to the $1p1h$ subspace and coincides with the vector in Eq.~(\ref{erpabe}).
The vector $|\,\zeta^{\,\nu}_{\vphd} \rangle$ belongs to the subspace
of complex configurations.

The matrix $\widehat{\Omega}^{\mbss{ERPA}}_{\vphd}$ can be represented
in the form
$\widehat{\Omega}^{\mbss{ERPA}}_{\vphd}=
{M}^{\mbss{ERPA}}_{\vphd}\mathfrak{S}^{\mbss{ERPA}}_{\vphd}$ where
\be
{M}^{\mbss{ERPA}}_{\vphd} =
\left(
\begin{array}{cc}
M^{\mbss{RPA}}_{\vphd} & 0 \\
0\hphantom{A} & M^{\mbss{C}}_{\vphd} \\
\end{array}
\right)
\label{merpa}
\ee
is the metric matrix,
$\mathfrak{S}^{\mbss{ERPA}}_{\vphd}$ is the stability matrix in the ERPA
which is defined in analogy to Eq.~(\ref{srpa}):
\be
\mathfrak{S}^{\mbss{ERPA}}_{\vphd} = {M}^{\mbss{ERPA}}_{\vphd}
\,\widehat{\Omega}^{\mbss{ERPA}}_{\vphd}.
\label{serpa}
\ee
Using Eqs. (\ref{oerpa3}), (\ref{orpa2}), (\ref{merpa}), and (\ref{serpa})
we obtain that for any complex vector $|Z\, \rangle$,
\be
|Z\, \rangle =
\left(
\begin{array}{c}
|\,z\, \rangle \vphantom{\frac{A}{B}} \\
|\,\zeta\, \rangle \vphantom{\frac{A}{B}} \\
\end{array}
\right),
\label{zext}
\ee
with arbitrary components $|\,z\, \rangle$ and $|\,\zeta\, \rangle$
the following equation is fulfilled
\bea
\langle\,Z\,|\,\mathfrak{S}^{\mbss{ERPA}}_{\vphd}|Z\, \rangle &=&
\langle\,z\,|\,\mathfrak{S}^{\mbss{RPA}}_{\vphd}|\,z\, \rangle +
\langle\,\zeta'|\,{\mathfrak{S}}^{\mbss{C}}_{\vphd}|\,\zeta'\, \rangle
\nonumber\\
&+& (\kappa-1)\,
\langle\,\zeta''|\,{\mathfrak{S}}^{\mbss{C}}_{\vphd}|\,\zeta''\, \rangle
\label{zserpaz}
\eea
where
\be
|\,\zeta'\, \rangle = |\,\zeta\, \rangle + |\,\zeta''\, \rangle,
\quad
|\,\zeta''\, \rangle = ({\mathfrak{S}}^{\mbss{C}}_{\vphd})^{-1}
{F}^{\dag}|\,z\, \rangle\,.
\label{ynsdef}
\ee
From Eq.~(\ref{zserpaz}) it follows that the expectation value
$\langle\,Z\,|\,\mathfrak{S}^{\mbss{ERPA}}_{\vphd}|Z\, \rangle
\geqslant 0$ for all $|Z\, \rangle$
if the RPA stability matrix $\mathfrak{S}^{\mbss{RPA}}_{\vphd}$
is positive semidefinite, the matrix ${\mathfrak{S}}^{\mbss{C}}_{\vphd}$
is positive definite, and $\kappa \geqslant 1$.
That is, under these conditions, the matrix
$\mathfrak{S}^{\mbss{ERPA}}_{\vphd}$ is positive semidefinite.
Note that the positive definiteness of the matrix
${\mathfrak{S}}^{\mbss{C}}_{\vphd}$ is ensured in the models considered
in Sec.~\ref{sect4}
and that  Eqs. (\ref{fhsym}) and (\ref{fsym1}) are not used in the proof
of this statement.
Since the matrices ${M}^{\mbss{ERPA}}_{\vphd}$ and
$\mathfrak{S}^{\mbss{ERPA}}_{\vphd}$ are Hermitian, in analogy to the case
of the RPA (see Sec.~\ref{sect2})
we conclude that all eigenvalues $\omega^{\vphu}_{\nu}$
in Eqs. (\ref{erpabe}) and (\ref{erpabe2}) are real if the subtraction
method ($\kappa=1$ in Eq.~(\ref{oerpa2})) is used. Without subtraction
($\kappa=0$) stability of the solutions of the ERPA equations is not
guaranteed.

Let us introduce the ERPA response function in the extended space
\be
\widehat{R}^{\mbss{ERPA}}_{\vphd}(\omega) = -
\bigl(\,\omega - \widehat{\Omega}^{\mbss{ERPA}}_{\vphd}\bigr)^{-1}
{M}^{\mbss{ERPA}}_{\vphd}\,.
\label{rfdef3}
\ee
The analysis of Sec.~\ref{sect3} is straightforwardly
generalized to the case
of the ERPA with subtraction. The orthonormalization condition for
the non-spurious eigenvectors $|\,Z^{\,\nu}_{\vphd}\rangle$ of the matrix
$\widehat{\Omega}^{\mbss{ERPA}}_{\vphd}$ in the extended space has
the form
\be
\langle \,Z^{\,\nu}_{\vphd}|\,{M}^{\mbss{ERPA}}_{\vphd}
|\,Z^{\,\nu'}_{\vphd}\rangle = \mbox{sgn}\,(\omega^{\vphu}_{\nu})\,
\delta^{\vphu}_{\nu,\,\nu'}
\label{znorm3}
\ee
which is analogous to Eq.~(\ref{zmz}).
In the $1p1h$ subspace from this condition and from Eqs.
(\ref{wdef1}), (\ref{oerpa3})--(\ref{merpa}) we obtain
\be
\langle \,z^{\,\nu}_{\vphd}|\,M^{\mbss{RPA}}_{\vphd}
- W^{D(\nu\nu')}_{\vphd}\,|\,z^{\,\nu'}_{\vphd}\rangle =
\mbox{sgn}\,(\omega^{\vphu}_{\nu})\,\delta^{\vphu}_{\nu,\,\nu'}
\label{znorm4}
\ee
where
\bea
W^{D(\nu\nu')}_{\vphd} &=&
\frac{W(\omega^{\vphu}_{\nu}) - W(\omega^{\vphu}_{\nu'})}
{\omega^{\vphu}_{\nu} - \omega^{\vphu}_{\nu'}}\,,
\quad \nu \ne \nu',
\label{wnn1}\\
W^{D(\nu\nu)}_{\vphd} &=&
\biggl(\frac{d\,W(\omega)}{d\,\omega}\biggr)_{\omega\,=\,\omega^{\vphu}_{\nu}}.
\label{wnn2}
\eea
From Eqs. (\ref{znorm3}) and (\ref{znorm4}) we see that, as in the RPA case,
the eigenvectors with positive eigenvalues in the ERPA have positive norms.

Using the known properties of the block matrices one can readily show that
$\widehat{R}^{\mbss{ERPA}}_{12,34}(\omega)=R^{\mbss{ERPA}}_{12,34}(\omega)$
where $\widehat{R}^{\mbss{ERPA}}_{12,34}(\omega)$ is the block of the matrix
$\widehat{R}^{\mbss{ERPA}}_{\vphd}(\omega)$ in the $1p1h$ subspace and
the matrix $R^{\mbss{ERPA}}_{\vphd}(\omega)$ is defined by
Eq.~(\ref{rfdef2}).
Then from the results obtained in Sec.~\ref{sect3} and from
the positive semidefiniteness of the stability matrix
$\mathfrak{S}^{\mbss{ERPA}}_{\vphd}$ at $\kappa=1$ it follows that
in the case, when the subtraction method is used,
the expansion of the type (\ref{rfexp3}), where
the matrices $a^{n}$ are Hermitian and positive semidefinite,
is valid for the response function $R^{\mbss{ERPA}}_{\vphd}(\omega)$.
Therefore, for the dynamic polarizability
\be
\Pi^{\mbss{ERPA}}_{\vphd}(\omega) = -
\langle\,Q\,|\,R^{\mbss{ERPA}}_{\vphd}(\omega)\,|\,Q\,\rangle
\label{perpadef}
\ee
the expansion of the type (\ref{piqrpa}) holds where the probabilities
$B^{\vphu}_{\,n}$ are real and non-negative.

Though the problem of the convergence is not generally resolved within
the framework of the subtraction method, one can see that
its use at least improves the situation.
This problem arises when the model configuration space is enlarged,
i.e. when $\Omega^{\vphu}_{c}$ in Eq.~(\ref{wdef2}) increases.
Let us denote $\bar{{W}}(\omega)={W}(\omega)-{W}(0)$.
From Eq.~(\ref{wdef2})
one obtains the following formal expansions
\be
{W}(\omega) = - \sum_{c,\;\sigma}
\,\frac{
|\,{F}^{\,c(\sigma)}_{\vphd}\rangle
\langle {F}^{\,c(\sigma)}_{\vphd}|}{\Omega^{\vphu}_{c}}\;
\sum_{m=0}^{\infty}
\biggl(\frac{\sigma\omega}{\Omega^{\vphu}_{c}}\biggr)^m,
\label{w0exp}
\ee
\be
\bar{{W}}(\omega) = - \sum_{c,\;\sigma}
\,\frac{
|\,{F}^{\,c(\sigma)}_{\vphd}\rangle
\langle {F}^{\,c(\sigma)}_{\vphd}|}{\Omega^{\vphu}_{c}}\;
\sum_{m=1}^{\infty}
\biggl(\frac{\sigma\omega}{\Omega^{\vphu}_{c}}\biggr)^m.
\label{w1exp}
\ee
The convergence is determined by the rate of decrease of the terms in
these expansions at $\,\Omega^{\vphu}_{c} \rightarrow \infty\,$.
The leading term in the expansion (\ref{w0exp}) is of order
$1/\Omega^{\vphu}_{c}\,$, while in the expansion (\ref{w1exp}) this term
is of order $1/\Omega^2_{c}$.
Thus, the use of the quantity $\bar{{W}}(\omega)$ instead of ${W}(\omega)$
in the subtraction method leads to the acceleration of the convergence.

\section{The case of a schematic model}
\label{sect7}

To illustrate the results of the previous sections consider a simple model
in which the space of $1p1h$ states is restricted to one
particle-hole ($ph$) pair with the single-particle energies
$\ve^{\vphu}_p$ and
$\ve^{\vphu}_h$ and with the matrix elements of the residual
interaction ${V}^{\vphu}_{ph,ph}={V}^{\vphu}_{hp,\,hp}$ and
${V}^{\vphu}_{ph,\,hp}={V}^{\vphu}_{hp,\,ph}$ which are
supposed to be real.
The space of the complex configurations is also restricted to one
state, so that index $c$ in Eq.~(\ref{wdef2}) takes only one value
and we put:
$\bigl|\,{F}^{c(+)}_{ph}\bigr|^2=\bigl|\,{F}^{c(-)}_{hp}\bigr|^2=g^2$,
$\;{F}^{c(-)}_{ph}={F}^{c(+)}_{hp}=0$.

Let us denote in accordance with usual notations of the RPA equations
\cite{RS80}:
\be
A = \ve^{\vphu}_p - \ve^{\vphu}_h
+ {V}^{\vphu}_{ph,ph}\,,
\quad
B = {V}^{\vphu}_{ph,\,hp}\,.
\label{abrpa}
\ee
Then we have
\be
\Omega^{\mbss{RPA}}_{\vphd} =
\left(
\begin{array}{rr}
A & B \\ -B & -A \\
\end{array}
\right),
\quad
\mathfrak{S}^{\mbss{RPA}}_{\vphd} =
\left(
\begin{array}{rr}
A & B \\ B & A \\
\end{array}
\right),
\label{matrpa}
\ee
\be
M^{\mbss{RPA}}_{\vphd} =
\left(
\begin{array}{rr}
1 & 0 \\ 0 & -1 \\
\end{array}
\right).
\label{mrpa22}
\ee
The eigenvalues of the matrix $\Omega^{\mbss{RPA}}_{\vphd}$
are $\pm\,\omega^{\vphu}_{\mbsu{RPA}}$ where
$\omega^{\vphu}_{\mbsu{RPA}}=\sqrt{A^2-B^2}$.
The eigenvalues of the matrix $\mathfrak{S}^{\mbss{RPA}}_{\vphd}$ are
$s^{\mbss{RPA}}_{\pm}=A \pm |B|$. So, the RPA stability condition
reads
\be
A \geqslant |B|\,.
\label{rpasc}
\ee
The ERPA matrix (\ref{oerpa2}) in this model has the form
\be
\Omega^{\mbss{ERPA}}_{\vphd}(\omega) =
\left(
\begin{array}{cc}
A^{\vphu}_{\kappa} + {C}(\omega) & B \\
-B & -A^{\vphu}_{\kappa} - {C}(-\omega) \\
\end{array}
\right)
\label{oerpasm}
\ee
where
\be
A^{\vphu}_{\kappa} = A
+ \frac{\kappa\,g^2}{\Omega^{\vphu}_c}\,,\qquad
{C}(\omega) = \frac{g^2}{\omega - \Omega^{\vphu}_c}\,.
\label{bphiq}
\ee
In what follows we suppose that Eq.~(\ref{rpasc}) is fulfilled and that
$\Omega^{\vphu}_c>\omega^{\vphu}_{\mbsu{RPA}}$. This corresponds
to the real conditions in the models described in Sec.~\ref{sect4}.

Let us introduce the following dimensionless quantities
\be
\beta=B/A\,,\quad
\gamma = g/\sqrt{A\Omega^{\vphu}_c}\,,\quad
\omega^{\vphu}_c = \Omega^{\vphu}_c/A\,,
\ee
\be
\bar{s}^{\mbss{RPA}}_{\pm}=s^{\mbss{RPA}}_{\pm}/A
= 1 \pm |\beta|\,.
\ee
Note that the parameter $\gamma$ determines the strength of the coupling
of the $ph$ pair with complex configuration.
Consider properties of the poles and residua of
the ERPA response function defined by Eq.~(\ref{rfdef2}).
Its poles coincide with roots of the secular equation
\be
\det\,\bigl(\Omega^{\mbss{ERPA}}_{\vphd}(\omega)
-\omega\bigr)=0
\label{deterpa}
\ee
which has four roots:
$\pm\,\omega^{\vphu}_{\tau}$ where $\tau=\pm 1$,
\bea
\omega^2_{\tau} &=& \frac{1}{2}\,
\bigl(\,{U}^2_{\kappa} + \tau D^2_{\kappa}\,\bigr)\,,
\label{rootk2}\\
{U}^2_{\kappa} &=& A^2\bigl[\,(1+\kappa \gamma^2)^2 + \omega^2_c - \beta^2
+ 2\omega^{\vphu}_c \gamma^2\,\bigr]\,,
\label{epsk2}\\
D^4_{\kappa} &=& {U}^4_{\kappa}
+ 4 A^4 \omega^2_c
\bigl(\,\beta^2 - [\,1+(\kappa - 1)\,\gamma^2\,]^{\,2}\,\bigr).
\label{delk2}
\eea
The values of $\omega^2_{\tau}$ are always real because
$D^4_{\kappa}\geqslant 0$ both at $\kappa=1$ and at $\kappa=0$.

Substituting Eqs. (\ref{mrpa22}) and (\ref{oerpasm})
into Eq.~(\ref{rfdef2}), we obtain
\be
R^{\mbss{ERPA}}_{\vphd}(\omega) = - \sum_{\tau,\,\sigma}
\frac{\sigma\,a^{\vphu}_{\tau,\,\sigma}}
{\omega - \sigma\,\omega^{\vphu}_{\tau}}
\label{rfsm}
\ee
where $\sigma=\pm 1$,
\be
a^{\vphu}_{\tau,\,\sigma} =
\frac{\tau\,(\omega^2_{\tau} - \Omega^2_c)}
{2\,\omega^{\vphu}_{\tau}D^2_{\kappa}}
\left(
\begin{array}{cc}
\tilde{A}^{\vphu}_{\kappa}(-\sigma\omega^{\vphu}_{\tau}) & -B \\
-B & \tilde{A}^{\vphu}_{\kappa}(\sigma\omega^{\vphu}_{\tau}) \\
\end{array}
\right),
\label{atsmatr}
\ee
$\tilde{A}^{\vphu}_{\kappa}(\omega) = A^{\vphu}_{\kappa}
+ {C}(\omega) - \omega$.
The residue matrices $a^{\vphu}_{\tau,\,\sigma}$ obey the
condition
\be
\mbox{Tr}\,\bigl(\,\bigl[\,M^{\mbss{RPA}}_{\vphd}
- W^D(\sigma\omega^{\vphu}_{\tau})\,\bigr]\,
a^{\vphu}_{\tau,\,\sigma}\bigr) = \sigma
\label{atsnc}
\ee
where
\be
W^D(\omega) = \frac{d}{d\,\omega}
\left(
\begin{array}{cc}
C(\omega) & 0 \\ 0 & C(-\omega) \\
\end{array}
\right).
\label{wd22}
\ee
In addition, we have
$\,\det\,(\,a^{\vphu}_{\tau,\,\sigma}) = 0$.
So, the $2\times 2$ matrix $a^{\vphu}_{\tau,\,\sigma}$
has only one non-zero eigenvalue
$\alpha^{\vphu}_{\tau}=\mbox{Tr}\,(\,a^{\vphu}_{\tau,\,\sigma})$
which does not depend on $\sigma$ and is determined by the formula
\be
\alpha^{\vphu}_{\tau} =
\frac{\tau A\bigl[\,(1+\kappa \gamma^2)\,
(\omega^2_{\tau} - \Omega^2_c) +\Omega^2_c\gamma^2\,\bigr]}
{\omega^{\vphu}_{\tau}D^2_{\kappa}}\,.
\label{evats}
\ee
The product $\alpha^{\vphu}_{\tau}\omega^{\vphu}_{\tau}$
is real for all $\gamma$, $\tau$, and $\kappa\,$. However, at $\kappa=0$
and $\tau=-1$ it changes sign at $\gamma^2=\gamma^2_0$ where
$\gamma^2_0 = 1 + \beta^2/(1+\omega^{\vphu}_{c})^2$ and
$\bar{s}^{\mbss{RPA}}_{-} < \gamma^2_0 < \bar{s}^{\mbss{RPA}}_{+}$.

In the limit $\gamma^2 \rightarrow 0$ we have
$\omega^2_{-} \rightarrow \omega^2_{\mbsu{RPA}}$,
$\omega^2_{+} \rightarrow \Omega^2_{c}$,
$\alpha^{\vphu}_{-}\,\omega^{\vphu}_{-} \rightarrow
\alpha^{\vphu}_{\mbsu{RPA}}\omega^{\vphu}_{\mbsu{RPA}}=A$,
$\,\alpha^{\vphu}_{+}\,\omega^{\vphu}_{+} \rightarrow 0$
both for $\kappa=1$ and for $\kappa=0$.

In the limit $\gamma^2 \rightarrow \infty$ we obtain
\bea
&&\kappa = 1 :
\;\;\omega^2_{-} \rightarrow 0,
\;\;\omega^2_{+} \rightarrow \infty,
\;\;\alpha^{\vphu}_{-}\,\omega^{\vphu}_{-} \rightarrow 0,
\;\;\alpha^{\vphu}_{+}\omega^{\vphu}_{+} \rightarrow \infty\,;
\nonumber\\
&&\kappa = 0 :
\;\;\omega^2_{\pm} \rightarrow \infty,
\;\;\;\alpha^{\vphu}_{\pm}\omega^{\vphu}_{\pm} \rightarrow \pm\,\infty\,.
\nonumber
\eea

From Eqs. (\ref{rootk2})--(\ref{delk2}) and (\ref{evats})
it follows that at $\kappa=1$ all $\omega^{\vphu}_{\tau}$
and $\alpha^{\vphu}_{\tau}$ are real and all
$\alpha^{\vphu}_{\tau}\omega^{\vphu}_{\tau}>0$.
In this case the normalization condition (\ref{znorm4}) is fulfilled
due to Eq.~(\ref{atsnc}).
The matrices $a^{\vphu}_{\tau,\,\sigma}$ are Hermitian and
positive semidefinite if we set $\omega^{\vphu}_{\tau}>0$
(that is always possible if $\omega^{\vphu}_{\tau}$ are real).

\begin{figure}[ht!]
\begin{center}
\includegraphics*[scale=0.5,angle=0]{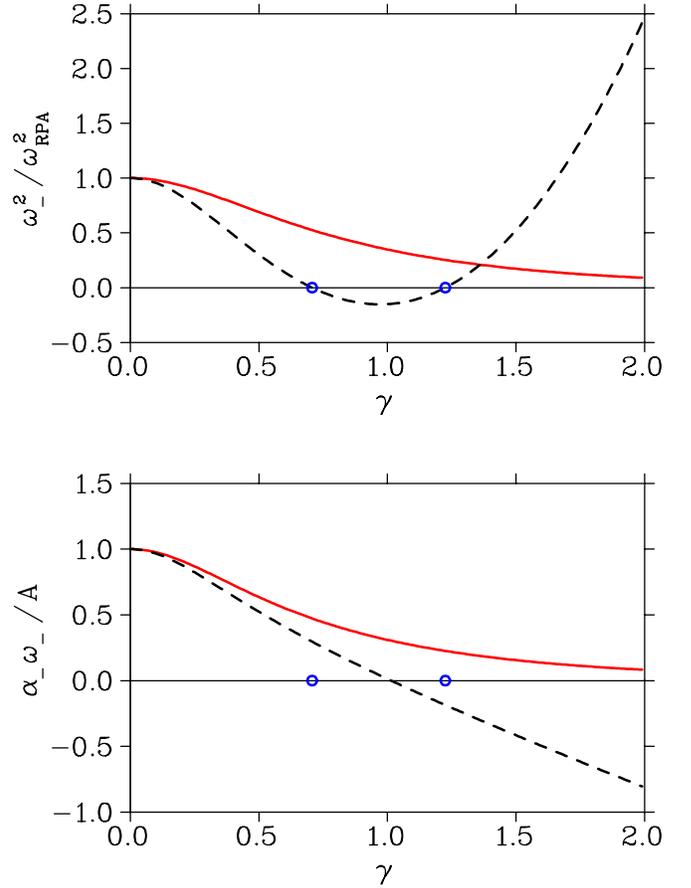}
\end{center}
\caption{\label{fig1}
Upper panel:
Dependence of the squared ERPA eigenvalue $\omega^2_{-}$
normalized to $\omega^2_{\mbsu{RPA}}$
on the parameter $\gamma$ determining the strength of the coupling
of the $ph$ pair with complex configuration.
The values of $\omega^2_{-}$ are calculated by means of
Eqs. (\ref{rootk2})--(\ref{delk2}) with $\beta = 0.5$ and
$\omega_{\mbss{$c$}} = 2$.
Solid line represents the ERPA results obtained with the use of
subtraction method ($\kappa=1$).
The dashed line represents the results without subtraction ($\kappa=0$).
The values of $(\bar{s}^{\mbss{RPA}}_{\pm})^{1/2}$ are indicated
by circles on the $\gamma$-axis.
Lower panel:
The same dependence for the product $\alpha_{_{\scs -}}\omega_{_{\scs -}}$
normalized to $\alpha_{_{\mbsu{RPA}}}\omega_{_{\mbsu{RPA}}}=A$
(see text for details).
}
\end{figure}

\begin{figure}
\begin{center}
\includegraphics*[scale=0.5,angle=0]{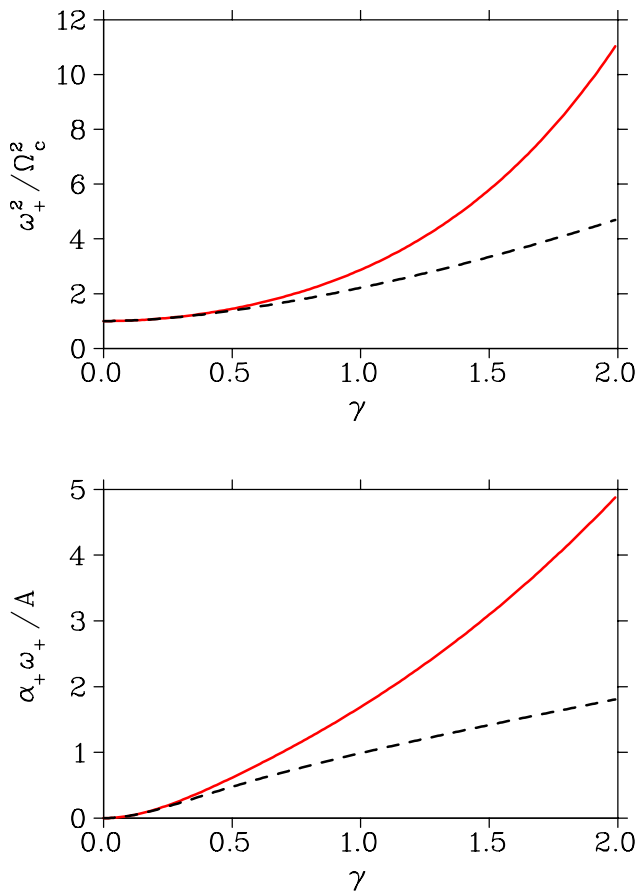}
\end{center}
\caption{\label{fig2}
The same as Fig.~\ref{fig1} for $\omega^2_{+}$ normalized to $\Omega^2_c$
(upper panel) and $\alpha_{_{\scs +}}\omega_{_{\scs +}}$ (lower panel).
}
\end{figure}

At $\kappa=0$ we have
\begin{itemize}
\item[(a)]
if $\gamma^2 < \bar{s}^{\mbss{RPA}}_{-}$, then
$\,\omega^{\vphu}_{\pm}\,$ and $\,\alpha^{\vphu}_{\pm}\,$ are real
and $\alpha^{\vphu}_{\pm}\omega^{\vphu}_{\pm}>0$;
\item[(b)]
if
$\bar{s}^{\mbss{RPA}}_{-} < \gamma^2 < \bar{s}^{\mbss{RPA}}_{+}$,
then $\omega^{\vphu}_{-}$ and $\alpha^{\vphu}_{-}$ are imaginary,
$\omega^{\vphu}_{+}$ and $\alpha^{\vphu}_{+}$ are real and
$\alpha^{\vphu}_{+}\omega^{\vphu}_{+}>0$;
\item[(c)]
if $\gamma^2 > \bar{s}^{\mbss{RPA}}_{+}$, then
$\omega^{\vphu}_{\pm}$ and $\alpha^{\vphu}_{\pm}$ are real,
$\alpha^{\vphu}_{+}\omega^{\vphu}_{+}>0$, but
$\alpha^{\vphu}_{-}\omega^{\vphu}_{-}<0$;
\item[(d)]
if $\gamma^2 = \bar{s}^{\mbss{RPA}}_{\pm}$, then
$\omega^{\vphu}_{+}$ and $\alpha^{\vphu}_{+}$ are real,
$\alpha^{\vphu}_{+}\omega^{\vphu}_{+}>0$,
$\omega^{\vphu}_{-} = 0$,
$\;\alpha^{\vphu}_{-}$ is indefinite.
\end{itemize}
These properties of the values $\omega^{\vphu}_{\pm}$ and
$\alpha^{\vphu}_{\pm}$ do not depend on the value of the parameter
$\omega^{\vphu}_c$ if $\Omega^{\vphu}_c>\omega^{\vphu}_{\mbsu{RPA}}$.

Dependence of the values $\omega^{2}_{\pm}$ and
$\alpha^{\vphu}_{\pm}\omega^{\vphu}_{\pm}$ on the parameter $\gamma$
at $\beta = 0.5$ and $\omega_{\mbss{$c$}} = 2$ is shown
in Figs. \ref{fig1} and \ref{fig2}.
Since $|\beta|<1$, the RPA stability condition (\ref{rpasc}) is fulfilled.
We see that in this case the ERPA solutions are also stable
and the eigenvalues of the ERPA residue matrices are non-negative
at all $\gamma$ (and $\omega^{\vphu}_{\pm}>0$)
if the subtraction method is used.
In the ERPA without subtraction
the lowest eigenvalue
$\omega^{\vphu}_{-}$ becomes imaginary in the finite region of
the values of $\gamma$ around the point $\gamma=1$.
Outside of this region, the values of $\omega^{\vphu}_{-}$
at $\kappa=0$ are real,
however at $\gamma^2 > \gamma^2_0 = 1 + \beta^2/(1+\omega^{\vphu}_{c})^2$
the product of $\omega^{\vphu}_{-}$ and the eigenvalue $\alpha^{\vphu}_{-}$
of the ERPA residue matrices $a^{\vphu}_{-,\,\pm}$ becomes negative.
Using Eq.~(\ref{atsnc}) we obtain that in this case the eigenvector with
the positive eigenvalue $\omega^{\vphu}_{-}$ will have the negative norm.
Therefore, the condition (\ref{znorm4}) is violated.
In addition, in the region $\gamma^2 > \bar{s}^{\mbss{RPA}}_{+}$, where
$\alpha^{\vphu}_{-}\omega^{\vphu}_{-}<0$, the matrices $a^{\vphu}_{-,\,\pm}\,$,
though being Hermitian,
become negative semidefinite at $\omega^{\vphu}_{-}>0$
(positive semidefinite at $\omega^{\vphu}_{-}<0$)
that leads to the problem of the ``negative transition probabilities''
(or of the ``negative energies'') as was explained in Sec.~\ref{sect3}.

From Eqs. (\ref{bphiq}) it follows that the subtraction effectively
introduces additional repulsion into the matrix elements
$A^{\vphu}_{ph,ph}$ and $A^{\vphu}_{hp,hp}$
of the matrix $\Omega^{\mbss{RPA}}_{\vphd}$.
As a result,
$\omega^2_{-}(\kappa=1)>\omega^2_{-}(\kappa=0)$
at least at $\gamma^2<\bar{s}^{\mbss{RPA}}_{-}$.
Nevertheless, as can be seen from Fig. \ref{fig1},
$\omega^2_{-} < \omega^2_{\mbsu{RPA}}$
for all $\gamma^2 > 0$ and $\kappa=1$ due to the attractive effect
of the dynamic part of the interaction ${C}(\omega)$ at
$\omega < \Omega^{\vphu}_c\,$.

\section{Conclusion}
\label{sect8}

In the paper the problem of stability of solutions in the extended
RPA (ERPA) theories is considered.
The extension of the RPA implies enlarging the configuration space
by taking into account more complex configurations in addition to
the $1p1h$ states included in the RPA.
The analysis of stability is based on the famous
Thouless theorem proved in the case of the self-consistent RPA
and on the response function formalism which enables one
to study this problem in more detail.
Two cases are considered: the ERPA with and without the subtraction method.
This method was suggested previously to avoid double counting
in the self-consistent ERPA approaches based on the density functional
theory with phenomenologically fitted energy density functionals.
Justification of the subtraction method is provided by the dielectric
theorem which associates the static polarizability calculated within the
self-consistent RPA with the exact inverse energy-weighted moment
of the strength distribution including contributions of all configurations.
The subtraction method ensures the equality of the RPA and of the ERPA
static polarizabilities and, consequently, equality of the respective
inverse energy-weighted moments.

It is proved that the stability matrix in the ERPA theories with subtraction
is positive semidefinite if the RPA stability matrix possesses this property.
This ensures stability of solutions of the ERPA eigenvalue equations,
positiveness of the norms of the eigenvectors with positive eigenvalues,
and non-negativeness of the respective transition probabilities.
In the ERPA without subtraction these properties of the solutions
are not guaranteed.
In addition, it is shown that the subtraction method leads to the
acceleration of the convergence in the ERPA though this problem is not
generally resolved within the framework of this method.

The example of the schematic model is used to analyze
dependence of the solutions of the ERPA equations
on the effective parameter determining the strength of the coupling
of a single particle-hole pair with a single complex configuration.
It is demonstrated that,
if the values of this parameter are sufficiently large,
the ERPA without subtraction leads to the imaginary
solutions of the respective eigenvalue equation and to
the problem of the ``negative transition probabilities'' or of
the ``negative energies''.
As in the general case, these problems do not arise when
the subtraction method is applied.

\begin{acknowledgements}
The author acknowledges financial support from the St. Petersburg State
University under Grant No. 11.38.648.2013.
\end{acknowledgements}

\end{document}